# Polydopamine-Coated TiO$_2$ Nanotubes for Selective Photocatalytic Oxidation of Benzyl Alcohol to Benzaldehyde under Visible Light

by Jyotsna Tripathy,[a] Gabriel Loget,[a,c] Marco Altomare,[a] and Patrik Schmuki[a,b,]*

[a] Prof. Dr. Patrik Schmuki, Dr. Jyotsna Tripathy, Dr. Gabriel Loget, and Dr. Marco Altomare

Department of Materials Science and Engineering, WW4-LKO, University of Erlangen-Nuremberg, Martensstrasse 7, D-91058 Erlangen, Germany.

* Corresponding author E-mail: schmuki@ww.uni-erlangen.de

[b] Prof. Dr. Patrik Schmuki

Chemistry Department, Faculty of Sciences, King Abdulaziz University, 80203 Jeddah, Saudi Arabia Kingdom

[c] Dr. G. Loget

Institut des Sciences Chimiques de Rennes, UMR 6226 CNRS/Université de Rennes 1, MaCSE, Campus de Beaulieu, 35042 Rennes Cedex, France

**A B S T R A C T**

TiO$_2$ nanotube arrays grown by anodization were coated with thin layers of polydopamine as visible light sensitizer. The PDA-coated TiO$_2$ scaffolds were used as photocatalyst for selective oxidation of benzyl alcohol under monochromatic irradiation at 473 nm. Benzaldehyde was selectively formed and no by-products could be detected. A maximized reaction yield was obtained in O$_2$-saturated acetonitrile. A mechanism is proposed that implies firstly the charge carrier generation in polydopamine as a consequence of visible light absorption. Secondly, photo-promoted electrons are injected in TiO$_2$ conduction band, and subsequently transferred to dissolved O$_2$ to form O$_2^{\bullet-}$ radicals. These radicals react with benzyl alcohol and lead to its selective dehydrogenation oxidation towards benzaldehyde.

**Introduction**

Since 1970s, photocatalysis has been largely investigated and applied to several environmental processes. Among these the most addressed are by far the degradation of pollutants (organics, NO$x$, pesticides, *etc.*) and the generation of energy vectors (*e.g.*, $H_2$, $CH_4$ and mixtures of hydrocarbons).[1-3]

Different semiconductors are being studied, such as metal oxides, sulfides, nitrides and oxonitrides. Among these, titanium dioxide ($TiO_2$) has outstanding potentiality for driving efficient photocatalysis (particularly the anatase polymorph) owing to its (photo)chemical stability and band-edge levels. $TiO_2$ absorbs UV light and effectively generates highly oxidative species that can, for instance, mineralize various pollutants.[2,3]

Besides, at the cutting-edge of research on photocatalysis, the synthesis of valuable organics is gaining large attention. The charge carriers formed when irradiating the semiconductor can be used in particular conditions to catalyze the conversion of a compound into a specific product[4-6] (*e.g.*, an alcohol into a certain carbonyl compound).[7,8] This process implies a fine chemistry (*e.g.*, a "mild" oxidation instead of a photocatalytic mineralization) as here one's focus is on synthesizing a specific molecule with high yield and selectivity, rather than indiscriminately oxidizing organic compounds to $CO_2$.

Thus, a catalytic synthesis that in a classic approach may imply severe experimental conditions (high pressure and/or high temperature),[9] the use of hazardous and costly catalysts and co-catalysts (typically oxygen donors such as chromate and permanganate,[9] or noble metals, *e.g.*, Ru),[10] and the production of large amounts of wastes,[9] can be replaced by a photocatalytic reaction that selectively delivers the desired product at room temperature and atmospheric pressure.

Even more interesting is this approach when driven by visible (solar) light.[11-17] However $TiO_2$ has a band gap of *ca*. 3-3.2 eV that corresponds to a light absorption threshold of *ca*. 390-410 nm. To enable visible light absorption and consequent charge carrier generation different strategies can be used to modify the photocatalyst such as surface modification with visible light absorbers,[18] non-metal doping,[19] incorporation of carbonaceous nanomaterials (activated carbon, carbon nanotubes, fullerenes, graphene *etc.*),[20] coupling with other semiconductors,[21] and deposition of plasmonic metals.[22,23]

Among these the sensitization of titania with a visible light absorber has been widely investigated especially in the field of dye-sensitized solar cells.[24] In general this concept is based on the absorption of visible light by the sensitizer which then in its excited state injects electrons from its LUMO (or *CB* if the sensitizer is a semiconductor) into the conduction band of $TiO_2$ (key is a proper alignment of the energetic levels).

In this work we fabricate $TiO_2$ nanotube (NT) arrays and coat them with thin films of polydopamine (PDA, a biomimetic polymer)[25,26] to achieve visible light sensitization of the tube scaffolds.[27,28] The PDA-coated tubes are then used for the oxidation of benzyl alcohol under monochromatic irradiation at 473 nm. In these irradiation conditions only PDA is photo-active (*i.e.*, absorbs light and generates charge carriers) while the oxide film is used, owing to its ordered porous structure, as semiconductive scaffold that allows for electron transfer from PDA to $TiO_2$ conduction band. We demonstrate that benzyl alcohol can be selectively converted to benzaldehyde in the presence of $O_2$ and propose a reaction mechanism that involves the dehydrogenation oxidation of benzyl alcohol ascribed to $O_2^{\bullet-}$ radicals.

**Results and Discussion**

$TiO_2$ nanotube arrays were selected as photocatalytic platform owing to their facile fabrication and high surface area which is key for efficient catalysis.[29,30] Nanotubes were preferred over nanoparticulate titania since they grow *i)* directly on a substrate and the separation of the catalyst from the reaction phase is facile, and *ii)* in highly ordered fashion allowing for controlled deposition of PDA.[28,29]

Arrays of TiO$_2$ NTs as the one shown in Fig 1a were grown by anodization of Ti foils at 60 V for 4 min in an ethylene glycol-based electrolyte (0.135 M NH$_4$F, 3 vol % H$_2$O). The tube films showed average thickness of ~ 4 µm and individual tube inner diameter of ~ 50 nm.

As reported in the literature as-grown nanotubes are amorphous.[31] Their crystallization into anatase TiO$_2$ was thus obtained by annealing the layers in air, for 1 h at 450°C.[31] Afterwards, the crystalline nanotube films were subjected to PDA deposition.

PDA is a biomimetic eumelanin polymer that not only exhibits unique adhesive properties but also delivers interesting functionalities when deposited as thin film coatings.[26,32-35] In particular, PDA behaves as a semiconductor with tunable electronic properties that depend on its physico-chemical state and hydration.[26,36] Owing to these features PDA is an attractive platform for the fabrication of optoelectronic[37,38] and photo-electrochemical devices.[27,28] Nam *et al.* and some of us recently reported on the coating of anatase TiO$_2$ nanoparticles[27] and TiO$_2$ NTs,[28] respectively, with PDA layers. The optical and photo-electrochemical properties of PDA deposited on these surfaces were investigated and the results indicated that PDA is able to effectively sensitize TiO$_2$ photo-anodes.[27,28]

Thus, to achieve visible light sensitization we coated the surface of the tube layers with PDA by using a procedure developed by Lee *et al.*[25] The NT films were immersed in a slightly alkaline dopamine solution for different times (previous studies illustrate how the thickness of the PDA coating can be finely tuned by adjusting the deposition time).[28] As shown by the SEM images in Fig 1b,c, a ~ 6 nm-thick layer of PDA was deposited onto the inner wall of the nanotubes after 24 h-long deposition, this confirming that the "mouth" of the NTs are open and allow access of the oligomers for PDA polymerization.

This observation is in good agreement with XPS results of uncoated and PDA-coated TiO$_2$ surfaces (Fig 1d), which show a decrease of the Ti and O peak intensity, a disappearing of the F peak and the appearance of the C and N peaks with increasing the PDA coating time. We also observed that the color of the anodic surface changes after PDA deposition and turned darker for longer coating times (shown in the optical pictures in Fig. 1e-g) owing to increased visible light absorbance of thicker PDA films.[39]

The PDA-coated $TiO_2$ NTs were tested as photocatalysts for the oxidation of benzyl alcohol under illumination at 473 nm (more details in the experimental section). Preliminary experiments were carried out using arrays of unmodified $TiO_2$ nanotubes or a Ti metal surface coated with PDA (irradiation time 2 h, coating time 3 h). In any case only benzyl alcohol was detected in the liquid phase by GC-MS, indicating that no photocatalytic reaction could take place (Scheme 1).

These results seem to be in contrast to what reported by Higashimoto *et al.*,[12] that is that benzyl alcohol undergoes selective oxidation towards benzaldehyde on bare titania under visible light illumination. Higashimoto and co-workers propose a reaction mechanism based on the adsorption of alcohol molecules onto $TiO_2$, which leads to a red-shift of titania light absorption features. Consequently, visible light (mainly in the 400-450 nm range) is absorbed and generates charge carriers that lead to benzyl alcohol selective oxidation. However the same authors reported in a follow-up study[13] that fluorinated $TiO_2$ does not show any visible light photo-activity. As $TiO_2$ NTs formed by anodization typically show some fluoride uptake from the electrolyte, we assume that the absence of photo-activity of pristine tubes is ascribed to a partial fluorination of the tube surface (the presence of fluorides on the annealed tubes was confirmed by XPS data – see Fig. 1d). Another possible cause is the relatively low power (*i.e.*, *ca.* 20 mW) of the monochromatic light used in the present work compared to the studies of Higashimoto *et al.*.[12]

When the reaction was performed with PDA-coated NTs, benzaldehyde was produced and it was the only reaction product (reaction selectivity of 100%) regardless of which solvent was used.

A first advantage of depositing PDA on the $TiO_2$ tubes is that the hybrid material (PDA/$TiO_2$) exhibits broad absorption in the visible range[26,27] (attributable to the intrinsic visible light absorption of PDA alone)[40] instead of a red-shifted absorption threshold as observed for alcohol-TiO$_2$ complexe.[12] Indeed, a broad band absorption ascribed to the use of a sensitizer is preferable over a red-shift of titania absorption features as in principle more visible light could be harvested with the former strategy. Secondly, the absorption features of PDA/TiO$_2$ allow for specific excitation of PDA only, that is, when illuminating with laser light at 473 nm only PDA absorbs and generate carriers while the TiO$_2$ scaffold is not photo-excited. This aspect is essential for carrying out selective oxidation and is discussed below when illustrating the proposed reaction mechanism.

As shown in Fig. 2, we found that the reaction solvent played a significant role over the benzaldehyde production rate. The formation of benzaldehyde was relatively small when the reaction was carried out in water (~ 1.8 μmol / 2 h), while was slightly larger in pure benzyl alcohol (*ca.* 2.2 μmol / 2 h in solvent-free conditions). A further significant improvement was observed in acetonitrile (~ 3.5 μmol / 2 h) although the concentration of the reactant (*i.e.*, benzyl alcohol) is clearly much smaller than in solvent-free conditions.

A possible explanation for this may relate to the polarity of the solvent and its ability in dissolving oxygen. In particular the solubility of O$_2$ in acetonitrile is higher than in water[41,42] and even larger amounts of benzaldehyde (~ 4.9 μmol / 2 h) were measured when pure O$_2$ was bubbled through the reaction phase. This indicates that O$_2$ plays a crucial role in the selective oxidation reaction. Nevertheless it is reported that in heterogeneous photocatalysis the use of acetonitrile as reaction solvent significantly increases (in comparison to water) the concentration of reactant within the surface solution monolayer[2] and this leads to higher benzaldehyde yields since more alcohol molecules are readily available to react (the reaction is less affected by diffusion in liquid phase).

Noteworthy, when running photocatalytic experiments only in the presence of pure acetonitrile, no reaction product could be detected and thus the solvent was considered to be stable. Also, the photocatalytic reaction yielded only benzaldehyde and did not proceed further, *e.g.*, towards benzoic acid which is otherwise one of the most common by-products along with other condensation compounds).[10]

As anticipated, the 473 nm light does not induce band gap excitation of the oxide so that holes in the $TiO_2$ valence band could not form. These holes would lead to strong (*i.e.*, unselective) oxidation of the organic molecules to $CO_2$.[2] The absorption of visible light exclusively by PDA is therefore a key (Scheme 2): after light absorption and charge separation in PDA, electrons are injected into the *CB* of $TiO_2$. Then the electrons can be transferred to $O_2$ molecules and form $O_2^{\bullet-}$ radicals that in turn enable a selective dehydrogenation oxidation of benzyl alcohol into benzaldehyde.[28,43]

As mentioned above, the solubility of $O_2$ is higher in acetonitrile than in benzyl alcohol or water.[41,42] Thus, the beneficial effect of using acetonitrile (instead of other solvents) is related to the amount of dissolved $O_2$, that is, the larger the amount of $O_2$ available in the reaction phase the larger the benzaldehyde production. This is confirmed also by the larger production of benzaldehyde observed when bubbling $O_2$ through the reaction phase. In other words, $O_2$-saturated acetonitrile provides a reaction phase with maximized $O_2$ supply for the photocatalytic reaction not to be kinetically hindered by a lack of $O_2^{\bullet-}$ radicals.

The effect of PDA thickness on the photocatalytic reaction was also investigated by testing PDA-coated NTs prepared by different coating times (data are compiled in Fig. 3). The yield of benzaldehyde was found to increase with increasing the PDA film thickness until a *maximum* for a coating time of 1 h (~ 7 μmol / 2 h) and then decreased for longer deposition times.

It is known that tuning the coating time allows for direct control over the PDA thickness.[26,28] Therefore the photocatalytic results in Fig. 3 indicate that the PDA film should be thick enough to grant efficient visible light sensitization for a maximized light absorption and consequent charge carrier generation. However, too thick PDA layers may hinder the transfer of electrons to $O_2$ molecules to form $O_2^{\bullet-}$ radicals. In fact PDA typically shows relatively low conductivity (in spite of its semiconductive properties) and passivates electrodes when deposited as thick films.[44]

24 h-long photocatalytic tests were carried out (data in Fig. 4) and it was observed that the amount of produced benzaldehyde increased over time. To verify the reproducibility of the results and the recyclability of the photocatalyst, these experiments were repeated by using the same photocatalyst (after washing with acetonitrile and drying in a $N_2$ stream) and a fresh $O_2$-saturated acetonitrile solution of benzyl alcohol. A benzaldehyde production of *ca*. 30 μmol / 24 h was always measured (inset in Fig. 4) confirming that no deterioration (*e.g.*, PDA degradation) of the photocatalyst could take place during extended reaction times.

**Conclusions**

We demonstrated that selective synthesis of benzaldehyde from benzyl alcohol is achieved on PDA-coated $TiO_2$ nanotubes under visible light irradiation. Key parameters for a maximized reaction yield are the solvent and the amount of dissolved $O_2$. In general these results show the potentiality of PDA-sensitized $TiO_2$ nanostructures as a platform for the selective synthesis of complex molecules driven by solar light, this representing a branch of photocatalysis to be more in depth explored.

**Experimental Section**

All chemicals were purchased from Sigma Aldrich and used as received. Prior to anodization, titanium foils (0.1 mm thick, 99.9% purity, Advent) were cut into 1.5 cm x 1.5 cm samples and degreased by sonication in acetone, ethanol, and deionized (DI) water. They were then rinsed with DI water and dried in a $N_2$ stream. $TiO_2$ NT layers were grown by anodizing the Ti foils in an ethylene glycol-based electrolyte (0.135 M $NH_4F$, 3 vol% M $H_2O$) at 60 V for 4 min. After anodization, the $TiO_2$ nanotubes films were rinsed with ethanol and dried in a $N_2$ stream. Then, the nanotubes were crystallized into anatase $TiO_2$ by annealing the as-formed layers in air for 1 h at 450°C.

For depositing the PDA coatings, the $TiO_2$ NTs (or a surface of metal titanium, in the case of the control sample) were taped in a beaker containing 30 mL of 10 mM Tris-buffer (pH adjusted to 8.6 with conc. HCl). 120 mg of dopamine were added and the solution was maintained under stirring for the desired time.

For the SEM images in Fig. 1b,c, the coating was prepared on as-formed tubes (*i.e.*, non-annealed tubes) in order to avoid the formation of double-walled tubular structures[29,31] that could interfere when assessing the thickness of the PDA coating. Conversely, for the preparation of the photocatalysts, the coating was performed on annealed anodic surfaces. After coating, the samples were rinsed with Milli-Q water and dried in a $N_2$ stream.

For morphological characterization of the samples, a field-emission scanning electron microscope (FE-SEM, Hitachi S4800) was employed. X-ray photoelectron spectroscopy (XPS, PHI 5600, US) was used to characterize the chemical composition of the samples, and to verify the effective deposition of PDA.

For photocatalytic reactions, the NT layers (with a nominal surface of 0.78 $cm^2$, and supported onto the Ti foil) were immersed in the liquid phase in a quartz cuvette and then exposed to a 473 nm laser source (MLB 473, 20 mW) under stirring for the desired time. For the experiments without solvent, 5 mL of benzyl alcohol were used. For the other runs, 3 mL of solvent and 2 mL of benzaldehyde were used. The solutions were analyzed after reaction and the reaction products were identified and quantified using gas chromatography – mass spectrometry (GCMS-QO2010SE, Shimadzu).

**Acknowledgements**

G.L. acknowledges the Alexander von Humboldt Foundation for the postdoctoral fellowship. The authors would also like to acknowledge the ERC, the DFG, and the Erlangen DFG cluster of excellence EAM for financial support.

**Figure Captions**

**Figure 1**

a-c) SEM images showing (a) the cross-sectional view of typical TiO$_2$ NT layers used in this work, and (b,c) the tube inner cavities coated by the PDA thin film (these images were taken along the height of the tube layer and the PDA thin film is indicated with white arrows); d) XPS spectra of uncoated (black) and PDA-coated TiO$_2$ NT arrays prepared by different coating times (red 1 h, blue 3 h and green 6 h); e-g) optical photographs of the (e) uncoated surface and of surfaces coated for (f) 3 h and (g) 6 h.

**Scheme 1**

Summary of the results obtained for the oxidation of benzyl alcohol at 473 nm. The selective conversion of benzyl alcohol into benzaldehyde was obtained only in the presence of PDA-coated TiO$_2$ nanotubes and under visible light irradiation. Conversely, the anodic TiO$_2$ nanotube arrays or the PDA film alone did not lead to any reaction.

**Figure 2**

Effect of the solvent on the benzaldehyde production rate. The amount of formed benzaldehyde is plotted against the different solvents (photocatalytic reaction time 2 hours, PDA coating time 3 h).

**Scheme 2**

Proposed reaction mechanism. After charge separation induced in PDA by visible light absorption, the conduction band electrons are transferred to the conduction band of TiO$_2$, and subsequently to adsorbed O$_2$ forming O$_2^{\bullet-}$ radicals. These radicals enable the selective dehydrogenation oxidation of benzyl alcohol into benzaldehyde formation.

**Figure 3**

Effect of the PDA coating thickness on the benzaldehyde production rate. The amount of formed benzaldehyde is plotted against the PDA coating time (reaction time 2 hours, reaction carried out in $O_2$-saturated acetonitrile).

**Figure 4**

Kinetic of benzaldehyde production. The amount of formed benzaldehyde is plotted against the photocatalytic reaction time (PDA coating time 1 h, reaction carried out in $O_2$-saturated acetonitrile). Inset: amount of benzaldehyde formed after three repeated reaction cycles (each reaction cycle lasted 24 h).

**Figure 1**

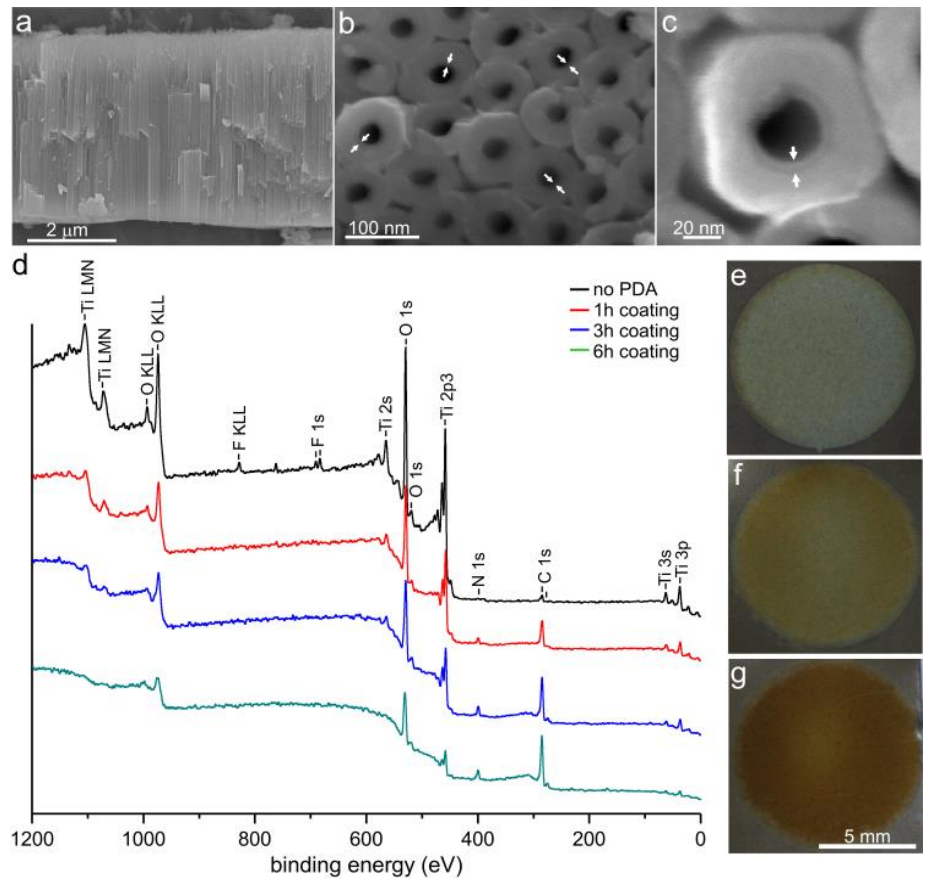

**Scheme 1**

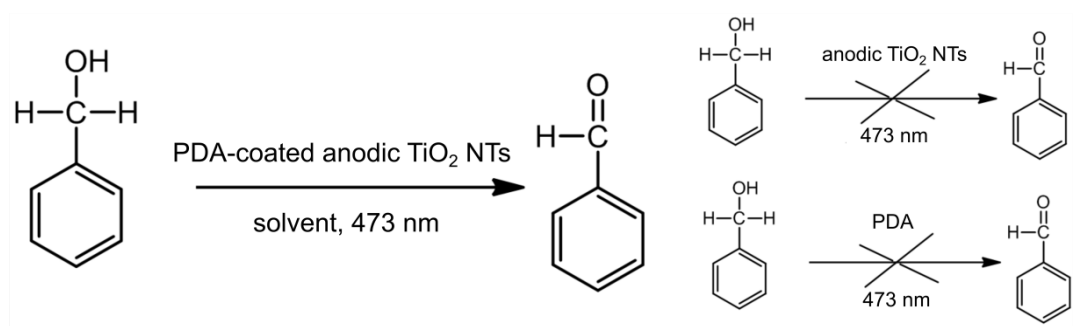

**Figure 2**

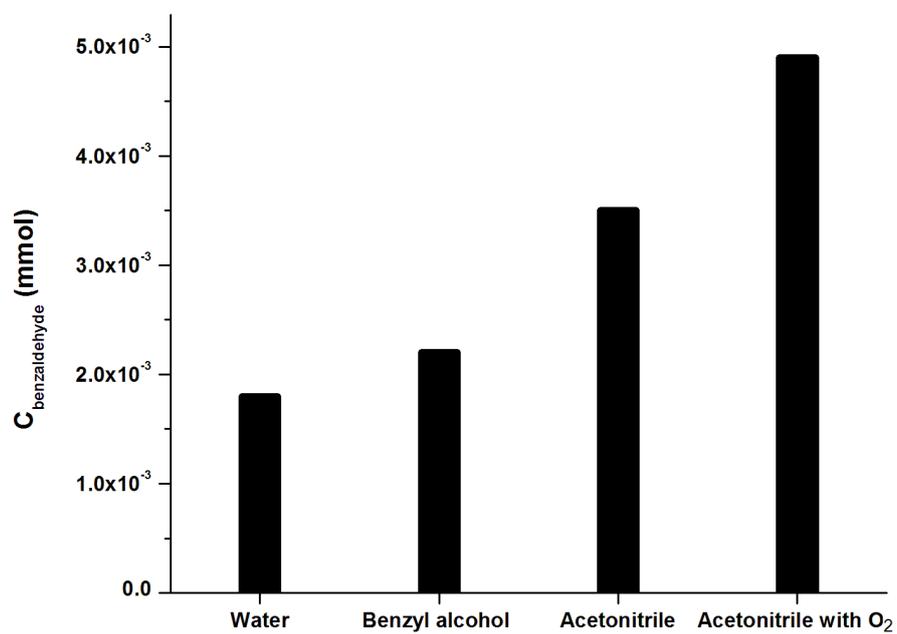

**Scheme 2**

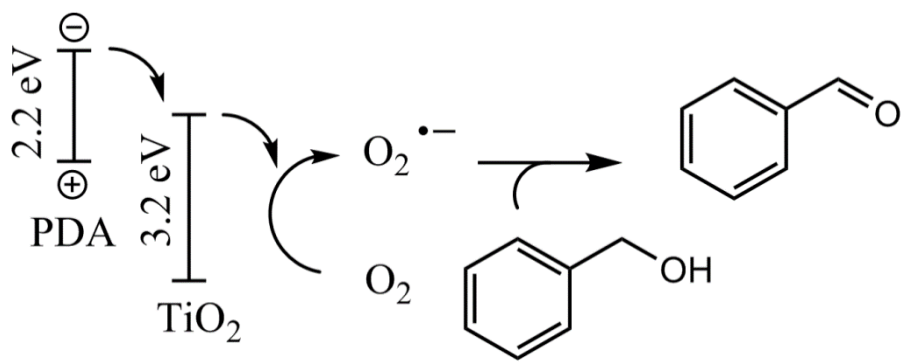

**Figure 3**

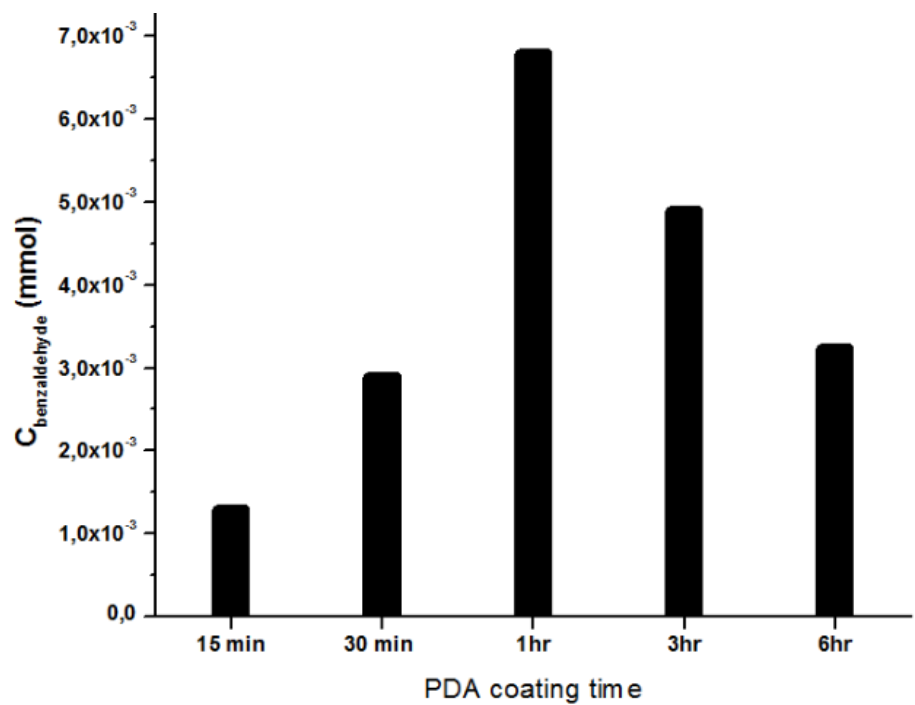

**Figure 4**

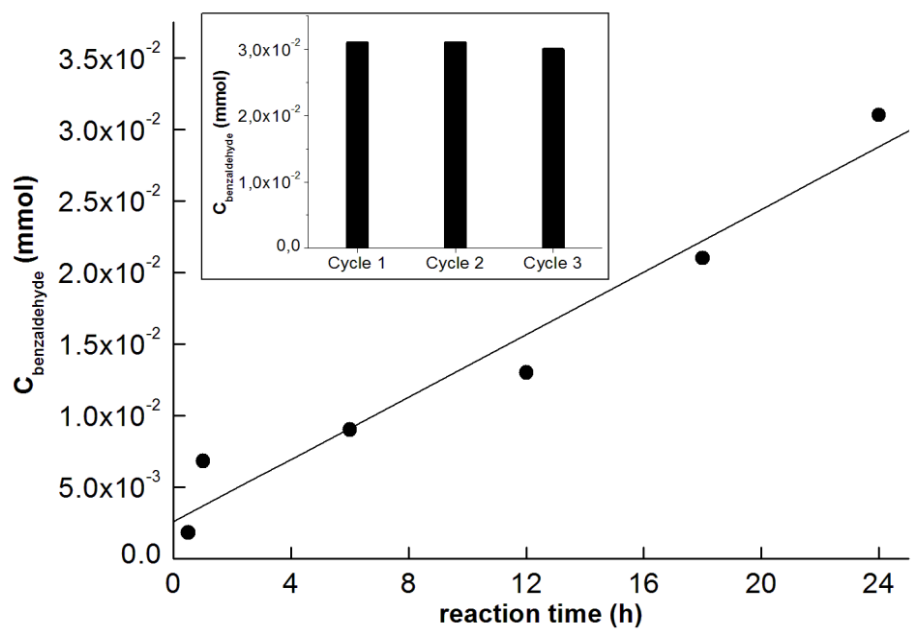

**Graphical abstract**

**Polydopamine-Coated TiO$_2$ Nanotubes for Selective Photocatalytic Oxidation of Benzyl Alcohol to Benzaldehyde under Visible Light** by Jyotsna Tripathy, Gabriel Loget, Marco Altomare, and Patrik Schmuki

Polydopamine-coated anodic TiO$_2$ nanotubes are used as photocatalyst for the selective oxidation of benzyl alcohol to benzaldehyde under visible light irradiation. A maximized reaction yield is obtained in O$_2$-saturated acetonitrile. The mechanism is proposed to proceed through O$_2^{\bullet-}$ radicals formation that in turn lead to the dehydrogenation oxidation of benzyl alcohol.

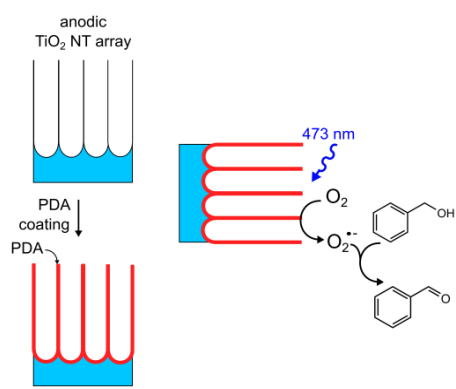